\newcommand{\BaCuSiO}{BaCu$_2$Si$_2$O$_7$}

\newcommand{\BaCuSiGexO}{BaCu$_2$(Si$_{1-x}$Ge$_x$)$_2$O$_7$}

\newcommand{\BaCuSiGeO}{BaCu$_2$(Si$_{0.5}$Ge$_{0.5}$)$_2$O$_7$}

\newcommand{\BaCuGeO}{BaCu$_2$Ge$_2$O$_7$}

\documentclass[twocolumn,prb,showpacs,preprintnumbers,amsmath,amssymb]{revtex4}
\usepackage{graphicx}
\usepackage{dcolumn}
\usepackage{bm}

\begin{document}

\title{Scaling of dynamic spin correlations in \BaCuSiGeO}

\author{A. Zheludev}
\affiliation{HFIR Center for Neutron Scattering, Oak Ridge
National Laboratory, Oak Ridge, Tennessee 37831-6393, USA.}

\author{T. Masuda}
\affiliation{International Graduate School of Arts and Sciences,
Yokohama City University, 22-2, Seto, Kanazawa-ku, Yokohama City,
Kanagawa, 236-0027, Japan.}

\author{G. Dhalenne}
\author{A. Revcolevschi}
\affiliation{Laboratoire de Physico-Chimie de l'Etat Solide,
Universite Paris- Sud, 91405 ORSAY Cedex, France.}

\author{C. Frost}
\author{T. Perring}
\affiliation{ISIS facility, Rutherford Appleton Laboratory,
Chilton, Didcot, Oxfordshire OX11 0QX, UK.}
\date{\today}

\begin{abstract}
The magnetic dynamic structure factor of the one-dimensional
$S=1/2$ chain system \BaCuSiGeO\ is studied in a wide range of
energy transfers and temperatures. Contrary to previous erroneous
reports [T. Masuda {\it et al.}, Phys. Rev. Lett. {\bf 93}, 077206
(2004)], the scaling properties observed in the range 0.5--25~meV
are found to be fully consistent with expectations for a Luttinger
spin liquid. At higher energies, a breakdown of scaling laws is
observed and attributed to lattice effects. The results are
complementary to those found in literature for other $S=1/2$-chain
compounds, such as KCuF$_3$ and Cu-benzoate.
\end{abstract}

\maketitle

\section{Introduction}

The two isostructural oxides \BaCuGeO\ and \BaCuSiO\ are
prototypical quasi-one dimensional (quasi-1D) quantum $S=1/2$
antiferromagnets (AF).\cite{Tsukada99} Previous extensive neutron
scattering experiments on \BaCuSiO\ were instrumental in
developing an understanding of the spin dynamics in weakly-coupled
$S=1/2$
chains.\cite{Zheludev2003-2,Zheludev2001-3,Zheludev2002-2,Zheludev2000-2}
More recently, solid solutions of type \BaCuSiGexO\ were
recognized as potential spin chain systems with random bond
strengths.\cite{Masuda2004} Bulk
measurements\cite{Yamada2001,Yamada2001-2,Masuda2004} and
preliminary neutron scattering studies\cite{Masuda2004} indicated
that the $x=0.5$ compound shows some very unusual scaling of the
static magnetic susceptibility and the dynamic spin correlation
function. It was proposed that, due to intrinsic quenched
structural disorder, \BaCuSiGeO\ is an experimental realization of
the much-studied Random Singlet (RS)
model.\cite{Dasgupta1980,Doty1992} Unfortunately, when followup
measurements were performed on larger and higher quality samples,
these conclusions were shown to be
erroneous.\cite{Masuda2006-erratum} In fact, in a wide energy
range the magnetic dynamic structure factor $S(q,\omega)$ follows
theoretical expectations for disorder-free spin chains. In the
present paper we report detailed temperature-dependent
measurements of magnetic excitations in \BaCuSiGeO\ obtained using
the new large single crystal samples. The focus is on scaling
relations for the spin correlation function. Convenient magnitudes
of magnetic interactions enable us to cover a wide dynamic range
$\hbar\omega/\kappa T\approx 1-100$. In this, our results are
 complementary to those previously obtained for two other
$S=1/2$ chain systems, namely Cu-benzoate\cite{Dender1997} and
KCuF$_3$ (Ref.~\onlinecite{Lake2005}), where $\hbar\omega/\kappa
T\approx 0.1-10$.

\begin{figure}
\includegraphics[width=8.7cm]{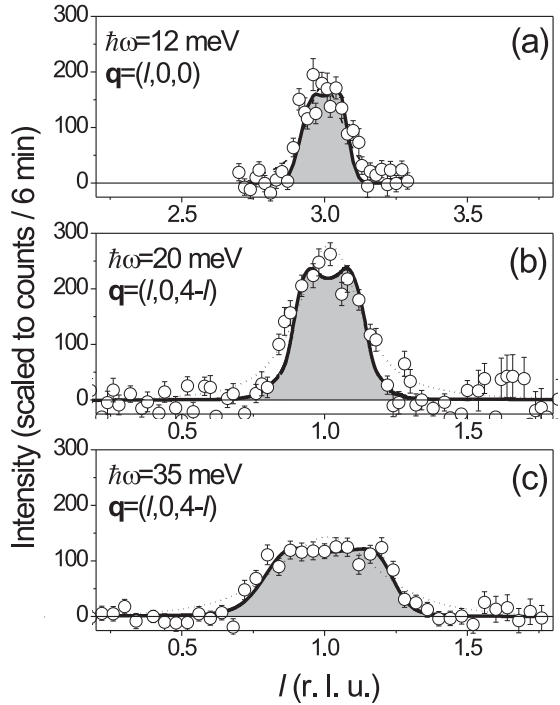}
\caption{Constant-$E$ scans measured in \BaCuSiGeO\ using Setup 1
(3-axis spectrometer) at $T=12$~K (symbols). Heavy solid lines and
shaded areas are fits using the M\"uller ansatz. Dashed lines are
Lorentzian fits. All fitting functions are convoluted with the
known experimental resolution. } \label{hb1data}
\end{figure}

\BaCuSiGexO\  crystallizes in an orthorhombic structure, space
group \textit{Pnma}. The lattice constants linearly depend on Ge
content and for $x=0.5$ are: $a = 6.917\,$\AA, $b = 13.28\,$\AA \,
and $c = 6.944\,$\AA. Magnetic properties are due to Cu$^{2+}$
ions that form weakly-coupled antiferromagnetic chains running
along the crystallographic $c$-axis. The in-chain exchange
constant is $J=24$~meV for the silicate ($x=0$) and $J=50$~meV for
the Ge-system ($x=1$).\cite{Tsukada99} The average coupling
strength $\langle J \rangle$ for intermediate concentrations can
be deduced from the position of the Bonner-Fisher susceptibility
maximum. It increases  linearly with
$x$,\cite{Yamada2001,Yamada2001-2} and for \BaCuSiGeO\ $\langle J
\rangle =37$~meV. The corresponding Des Cloizeau-Pearson (DCP)
zone-boundary energy of magnetic excitations is $\hbar
\omega_\mathrm{ZB}=\pi \langle J
\rangle/2=58$~meV.\cite{DesCloizeau1962} Nearest-neighbor
spin-spin separation within the chains is equal to $c/2$, so the
1D AF zone center $q_{\|}=\pi$ corresponds to $l=1$, where
$(h,k,l)$ denotes a vector in crystallographic reciprocal space.
Weak interactions between the chains result in long-range AF
ordering at $T_{\rm N }=$9.2~K in the silicate and $T_{\rm N
}=$10~K in the germanate.\cite{Tsukada99} However, for
intermediate concentrations, the transition temperature is
suppressed,\cite{Yamada2001,Yamada2001-2} and for \BaCuSiGeO, is
as low as $T_\mathrm{N}=0.7$~K.\cite{Masudaunpublished} Previous
theoretical and neutron scattering studies have shown that, in
weakly coupled spin chains, 3D effects are totally negligible at
energies $\hbar \omega \gtrsim 5 \kappa
T_\mathrm{N}$,\cite{Zheludev2003-2,Essler1997} so that, barring
any disorder-induced effects, \BaCuSiGeO\ can be expected to
behave like a good 1D system for $\hbar \omega \gtrsim 0.5$~meV.
Comparing this energy to $\hbar \omega_\mathrm{ZB}$, we see that
the material presents a conveniently broad energy range for
testing any predicted 1D scaling relations for
$S(\mathbf{q},\omega)$ in quantum $S=1/2$ chains.

\begin{figure}
\includegraphics[width=8.7cm]{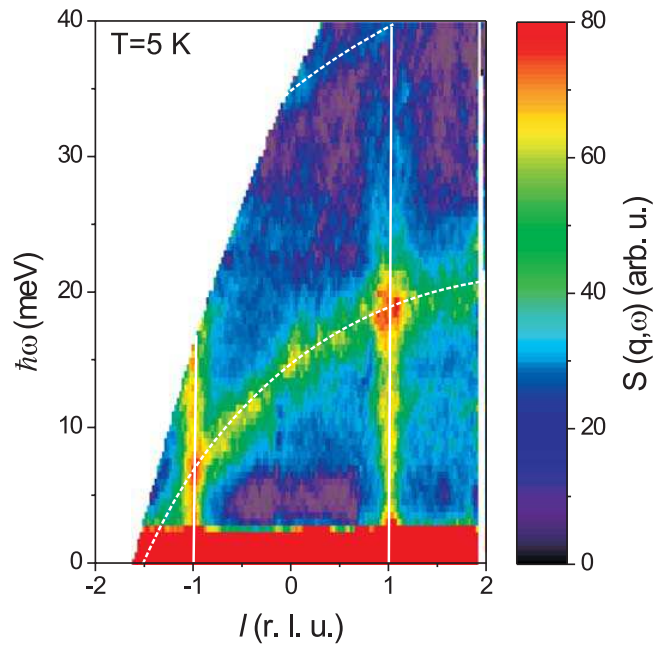}
\caption{(Color online) Time of flight spectrum collected in
\BaCuSiGeO\ at $T=5$~K using Setup 2, in projection onto the $c$
(chain) axis. Solid and dashed white lines are 1D
antiferromagnetic zone-centers for the two crystallographic
domains present in the sample, respectively. } \label{mapsdata1}
\end{figure}

\section{Experimental}
In the present study we used 6 \BaCuSiGeO\ single crystals of
total mass 15~g, grown using the floating zone method. Most
crystals were twinned in such a way that the $a$ and $c$ axes of
the two crystallographic domains were interchanged. Such a growth
habit is consistent with $a\approx c$, but complicates the
measurements, as it involves the chain axis. Individual crystals
were co-aligned to form a compound sample with a  total effective
mosaic of 1.7$^\circ$. Inelastic neutron scattering data were
collected in two separate series of experiments. Experimental
Setup 1 employed the HB1 thermal 3-axis spectrometer installed at
the High Flux isotope Reactor at ORNL. Final neutron energy was
fixed at 14.7~meV. A Pyrolitic Graphite (PG) monochromator was
used in conjunction with a PG filter installed after the sample.
The collimation setup was $48'-40'-40'-240'$. The sample was
mounted with the $(a,c)$ scattering plane horizontal, making
crystallographic wave vectors $(h,0,l)$ accessible for
measurements. Sample environment was a standard He-flow cryostat.
All data were corrected for the wavelength-dependent efficiency of
the beam monitor (due to higher-order beam contamination) and the
vertical focal length of the fixed-curvature monochromator. It
was, in part, an incorrect application of these corrections, that
led to erroneous conclusions in Ref.~\onlinecite{Masuda2004}. The
second series of measurements were performed on the MAPS
time-of-flight (TOF) chopper spectrometer installed at the ISIS
spallation neutron facility at RAL. Incident energy was fixed at
either 50~meV (Setup 2) or 100~meV (Setup 3). The $(a,c)$ plane of
the crystal was mounted parallel to the preferential scattering
plane of the instrument. All TOF data were integrated along the
$b^\ast$ crystallographic direction in the range $-1<k<1$. Sample
temperature was controlled between 5.5~K and 300~K using a
closed-cycle refrigerator.

\begin{figure*}
\includegraphics[width=17.4cm]{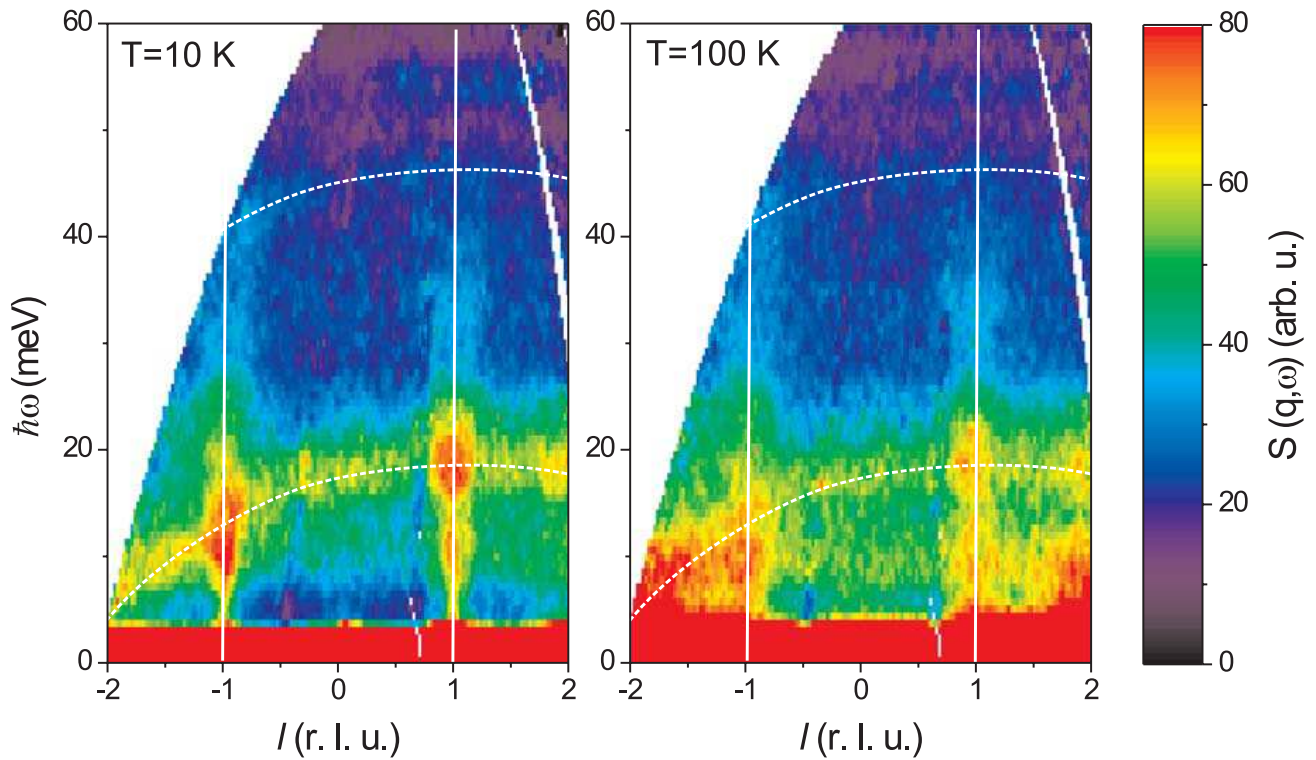}
\caption{(Color online) Time of flight spectrum collected in
\BaCuSiGeO\ at $T=5$~K using Setup 2, in projection onto the $c$
(chain) axis. Solid and dashed white lines are 1D
antiferromagnetic zone-centers for the two crystallographic
domains present in the sample, respectively. } \label{mapsdata2}
\end{figure*}

\section{Results and data analysis}

\subsection{3-axis data}
In 3-axis experiments the data were collected in constant-$E$
scans. Due to the presence of two sets of crystallographic
domains, any scattering at wave vector $(h,k,l)$ is observed
simultaneously with that at $(l,k,h)$. A steep dispersion of
magnetic excitations along the chains implies that, at low energy
transfers, all relevant scattering is concentrated near the 1D AF
zone-centers  $l$-odd. For this reason, scans performed at $l=0$
effectively pick up the signal from only one of the two domains.
While only half of the sample contributes to scattering, this
geometry is optimal for achieving a high wave vector resolution
along the chains. A typical constant-$E$ scan of this type
measured using Setup 1 at $\hbar \omega=12$~meV is shown in
Fig.~\ref{hb1data}a. An alternative approach is to restrict data
collection to $\mathbf{q}=(h\pm x,k,l\pm x)$, where $x$ is
arbitrary and $h$ and $l$ are integers. At these positions the
scattering in both crystallographic domains occurs at  wave
vectors that are equivalent in the 1D sense. The presence of two
domains still has to be carefully taken into account when
calculating the spectrometer resolution function. This second
approach is the only option of high energy transfers, where in
each domain the scattering is spread out over a wide $q$-range,
and necessarily interferes with that from the other domain.
Constant-$E$ scans collected in this mode are shown in
Fig.~\ref{hb1data}b,c. A constant background was subtracted from
all 3-axis scans shown.

\begin{figure}
\includegraphics[width=8.7cm]{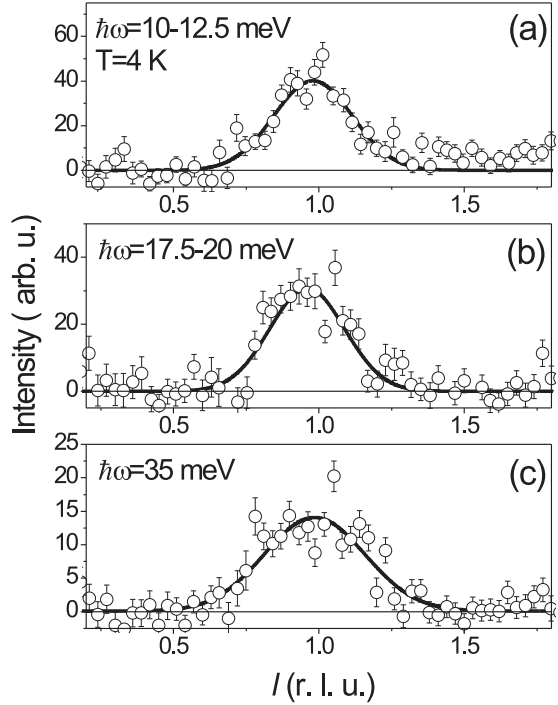}
\caption{Typical constant-$E$ cuts of the TOF data for \BaCuSiGeO\
collected using Setup 2 at $T=4$~K. Solid lines are Gaussian fits.
A flat background has been subtracted. } \label{mapscuts1}
\end{figure}

At low temperatures the magnetic dynamic structure factor of
disorder-free quantum $S=1/2$ chains can be accurately
approximated by the M\"uller Ansatz
\cite{Muller81,Dender1996,Lake2005,Zaliznyak2004,Tennant1995,Tennant1995-2,Zheludev2000-2}
(MA). For \BaCuSiGeO, in Ref.~\onlinecite{Masuda2004} it was
claimed that in constant-$E$ scans the magnetic scattering appears
broader than the MA expectation, and is more consistent with
Lorentzian line shapes. It was since realized that the elongated
Lorentzian-type ``tails'' are a result of background
contamination.\cite{Masuda2006-erratum} Analyzing the new data we
find that the MA peak profiles actually work for \BaCuSiGeO\
rather well. For each measured constant-$E$ scan the MA dynamic
structure factor was numerically folded with the known
spectrometer resolution function and fit to the experimental data.
The DCP spin wave velocity $v$ was fixed at $\pi\langle J\rangle
c/4=404$~meV\AA. The only variable parameter was an overall
scaling factor. Excellent fits were obtained at all energies
(solid lines and shaded areas in Fig.~\ref{hb1data}). Simulations
using Lorentzians convoluted with the experimental resolution
function are shown in dashed lines. Lorentzians provide
satisfactory fits only at low energy transfers, where the finer
spectral features are masked by the limited experimental
resolution anyway. At high energies deviations from Lorentzian
line shapes become obvious, and a flattening at the intensity
maximum resulting in typical ``top-hat'' shapes is clearly
observed (Fig.~\ref{hb1data}c).

\begin{figure}
\includegraphics[width=8.7cm]{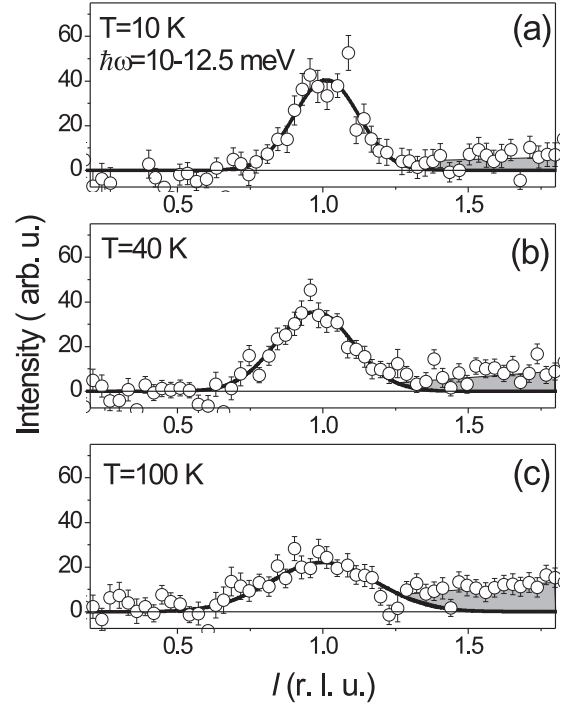}
\caption{Temperature evolution of a constant-$E$ cut measured for
\BaCuSiGeO\ using Setup 3. Solid lines are Gaussian fits. A flat
background has been subtracted. The shaded area is due to
scattering by phonons.} \label{mapscuts2}
\end{figure}

\subsection{TOF data}

Specifics of the TOF geometry make separating the signals
originating from the two crystallographic domains less
straightforward. Typical data collected using Setups 2 and 3 are
shown in Figs.~\ref{mapsdata1} and \ref{mapsdata2} in projection
onto the $c$-axis of one of the domains (domain A). Vertical
streaks of scattering at the 1D AF zone-centers $l=\pm 1$ for this
domain are clearly visible and highlighted by vertical solid
lines. However, they are intersected by the 1D AF zone-centers of
the 2nd domain (domain B), represented by the dashed curves. The
corresponding magnetic scattering from domain B appears as a
series of arcs. In order to separated it from domain-A scattering
one has to rely on certain assumptions and approximations. At
small energy transfers, where all the scattering is concentrated
in the vicinity of the 1D zone-centers, one can simply regard all
data collected sufficiently far from domain B 1D zone-centers as
belonging to domain A. Constant-energy cuts obtained from Setup 2
data using this assumption are shown in Fig.~\ref{mapscuts1}
(constant background subtracted). Compared to 3-axis data, the
experimental wave vector resolution of the TOF setup is
insufficient to resolve any fine structure in the scattering
profile.

\begin{figure}
\includegraphics[width=8.7cm]{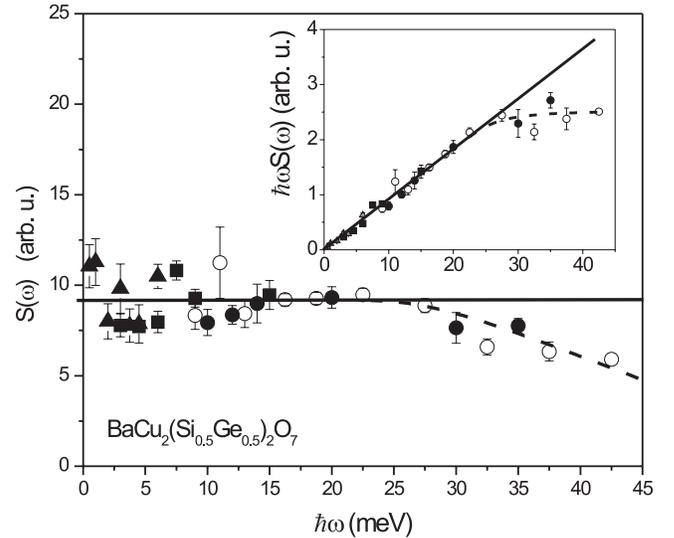}
\caption{Scaling of $S(\omega)$ measured in \BaCuSiGeO\ (symbols).
Solid circles: Setup 1 (3-axis), $T=12$~K. Open circles: setup 2
(TOF), $T=5.5$~K. Squares: Ref.\protect\onlinecite{Masuda2004},
re-analized thermal-neutron 3-axis data, $T=12$~K. Triangles:
Ref.\protect\onlinecite{Masuda2004}, cold-neutron 3-axis data,
$T=1.5$~K. Solid line: Luttinger spin liquid at $T\rightarrow 0$.
The dashed curves are guides for the eye. Inset: same plot for
$\hbar \omega S(\omega)$. } \label{lowT}
\end{figure}

An alternative approach is to select a geometry in which domain B
1D zone-centers cross that of domain A at an almost constant
energy, as in the case of $l=1$ in Fig.~\ref{mapsdata2}. For a
constant-$E$ cut extracted from such a data set, the magnetic
scattering by domain B will, to a good approximation, correspond
to a constant $l$-value. Since 1D magnetic scattering is
independent of $h$ and $k$, it will amount to a constant
background contribution. Constant-energy scans extracted from
Setup 3 data using this procedure are shown in
Fig.~\ref{mapscuts2}. A temperature-induced broadening of the
scattering profiles is clearly visible. Data analysis at high
temperatures is complicated by the increasing phonon contribution
to the background (Fig.~\ref{mapsdata2}b and shaded areas in
Fig.~\ref{mapscuts2}). For this reason, the measurements were
limited to $T\lesssim 100$~K.

\begin{figure}
\includegraphics[width=8.7cm]{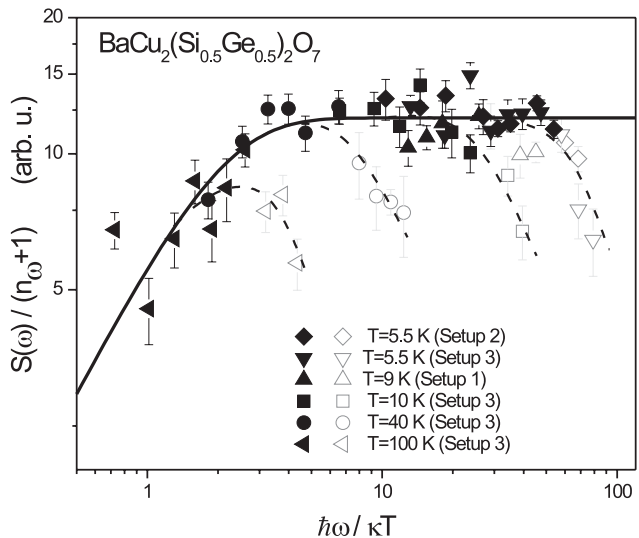}
\caption{Measured temperature scaling of $S(\omega)/(n_\omega+1)$
for \BaCuSiGeO. Open symbols correspond to energy transfers above
25~meV. The solid line is the approximate scaling function for a
Luttinger spin liquid.\protect\cite{Dender1996} Dashed lines are
guides for the eye and emphasize the breakdown of scaling at high
energies. } \label{allT}
\end{figure}

\subsection{Scaling quantities}
The central purpose of this study was investigating the energy and
temperature scaling of the spin correlations in \BaCuSiGeO. Much
of the existing experimental work concentrated on the scaling of
the dynamic structure measured at the 1D zone-center
$S(\pi,\omega)$.\cite{Dender1997,Lake2005} This quantity is
defined exclusively by long wavelength AF spin correlations. In
this regime quantum spin chains are equivalent to a Luttinger spin
liquid. The existing exact analytical predictions for this
model,\cite{Schulz86} make a comparison with theory
straightforward. The problem, however, is that measuring
$S(\pi,\omega)$ requires wave vector selectivity: the experimental
wave vector resolution $\delta q$ must be small compared to the
$q$-width of scattering at a particular energy transfer. The
latter can be estimated as $\hbar \omega/v$, where $v$ is the DCP
spin wave velocity. Thus, there will always be a lower limit on
energy transfer for the measurements, determined by the
experiemntal resolution: $\hbar \omega_\mathrm{min}\sim\delta q
v$. For our 3-axis setup, the practical lower limit is about
20~meV, and is still higher for the TOF experiments. A more robust
approach is to measure the $q$-integrated structure factor
$S_0(\omega)=\int S(q,\omega) d\omega$.\cite{Dender1997} In the
hydrodynamic limit the integral is to be taken over the entire
$q$-range, but, on a lattice, it should be limited to the vicinity
of the 1D AF zone-center. Taking the integral entirely eliminates
the effect of wave vector resolution. In our case of \BaCuSiGeO,
$S(\omega)$ was determined from the cross section fits to
individual 3-axis scans $q$-scans (solid lines in
Fig.~\ref{hb1data}). For the TOF data we employed empirical
Gaussian fits to constant-$E$ cuts, as shown in solid lines in
Figs.~\ref{mapscuts1} and \ref{mapscuts2}. The bulk of the
low-temperature data for $S(\omega)$ are shown in Fig.~\ref{lowT}
(circles). Re-analyzed data from Ref.~\onlinecite{Masuda2004} are
plotted as solid squares (thermal neutron  measurements) and
triangles (cold neutrons). The inset shows the same data scaled by
the energy transfer. All $S(\omega)$ data collected in this work
at different temperatures, after an appropriate scaling (see
below), are plotted in Fig.~\ref{allT}. Solid symbols are for
energy transfers $\hbar \omega<25$~meV, while open symbols
represent higher energies.

\section{Discussion}
The finite-temperature scaling for the dynamic structure factor of
a Luttinger spin liquid is given by:\cite{Schulz86}
\begin{eqnarray}
S(\tilde{q}, \omega)& \propto & (n_\omega+1) \frac{1}{T}
F\left(\frac{\hbar \omega}{\kappa T},\frac{v \tilde{q}}{\kappa
T}\right),\label{sqw}\\
F(x,y)& =
&\mathrm{Im}\left(\frac{\Gamma(\frac{1}{4}-i\frac{x-y}{4\pi})}{\Gamma(\frac{3}{4}-i\frac{x-y}{4\pi})}\frac{\Gamma(\frac{1}{4}-i\frac{x+y}{4\pi})}{\Gamma(\frac{3}{4}-i\frac{x+y}{4\pi})}\right),\label{Lut}
\end{eqnarray}
where $(n_\omega+1)=\left(1-e^{-\hbar \omega/(\kappa
T)}\right)^{-1}$. For an $S=1/2$ Heisenberg spin chain, this
scaling applies to AF spin correlations near a 1D AF zone-center
$q=q_0$, so that $\tilde{q}=q-q_0$. The integral over $\tilde{q}$
needed to derive the scaling for $S(\omega)$ cannot be easily
taken analytically. However, it can be very accurately
approximated by:\cite{Dender1996}
\begin{equation}
S(\omega)\sim \tanh\left (\frac{\hbar \omega}{2\kappa T} \right)
\label{scaling}.
\end{equation}
To a very good approximation, then, $S(\omega)=\mathrm{const}$ at
$\hbar \omega \gg \kappa T$. The same result can be obtained by
integrating the Muller Ansatz, which at small $\tilde{q}$ and
$\hbar\omega \ll \hbar \omega_\mathrm{ZB}$  coincides with the
$T\rightarrow 0$ asymptotic form for the scaling
function~\ref{Lut}.

The low-temperature data plotted in Fig.~\ref{lowT} show that in
\BaCuSiGeO, $S(\omega)$ is indeed practically energy independent
up to about 25~meV energy transfer (solid lines), {\it i. e.}, up
to about half of the zone-boundary energy $\hbar
\omega_\mathrm{ZB}$.  In this energy range the measured
temperature dependence of $S(\omega)$ is consistent with the
scaling form~\ref{scaling}, as shown in Fig.~\ref{allT}, solid
line. This plot, with a dynamic range for $\hbar\omega/\kappa T$
from 1 to 100, is complimentary to that shown in
Ref.~\onlinecite{Dender1996} for Cu-benzoate with
$\hbar\omega/\kappa T$ in the range of about 0.1 to 10. Above
25~meV, the measured $S(\omega)$ starts to decrease  and $\omega
S(\omega)$ deviates from linear behavior, even at low temperatures
(Fig.~\ref{lowT}, dashed lines). This breakdown of scaling is
observed at all temperatures, as shown by the open symbols and
dashed lines in Fig.~\ref{allT}.

The observed breakdown of scaling for $S(\omega)$ at about half
the zone-boundary energy is rather interesting, if not unexpected.
Work on KCuF$_3$ has shown that for $S(\pi,\omega)$ the scaling
relation~\ref{sqw} holds at much higher energies, up to about
140\% of $\hbar \omega_\mathrm{ZB}$.\cite{Lake2005} Of course, the
difference is due to the  fact that Eq.~\ref{sqw} is derived for
the continuum limit $|\tilde{q}|\ll 1$. Unlike $S(\pi,\omega)$,
$S(\omega)$ includes scattering at wave vectors
$-\hbar\omega/v<\tilde{q}<-\hbar\omega/v$, and thus rapidly
exceeds the domain of validity of Eq.~\ref{sqw} as $\hbar \omega /
v$ increases. Note that the existing $S(\omega)$ scaling data for
Cu-benzoate also extend to about half of $\hbar
\omega_\mathrm{ZB}$.

\section{Concluding remarks}
It is now clear that at energy transfers exceeding about 0.5~meV,
\BaCuSiGeO\ behaves as a prototypical disorder-free 1D quantum
$S=1/2$ system. In this sense, the claim of
Ref.~\onlinecite{Yamada2001} that \BaCuSiGexO\ compounds are
useful $S=1/2$ chain materials, with a continuously adjustable
exchange constant, is fully justified. On the other hand, the
suppression of the ordering temperature, the anomalous
low-temperature increase in the bulk
susceptibility,\cite{Masuda2004} recent ESR data
\cite{Smirnov_unpublished}, and preliminary neutron spin echo
measurements \cite{Ehlers_unpublished} clearly show that, at lower
energies (longer time scales), quenched disorder {\it does} become
relevant. Understanding the scaling of spin correlations in this
regime is a challenge for future studies.

\acknowledgements Research at ORNL was funded by the United States
Department of Energy, Office of Basic Energy Sciences- Materials
Science, under Contract No. DE-AC05-00OR22725 with UT-Battelle,
LLC.


\begin{thebibliography}{24}
\expandafter\ifx\csname
natexlab\endcsname\relax\def\natexlab#1{#1}\fi
\expandafter\ifx\csname bibnamefont\endcsname\relax
  \def\bibnamefont#1{#1}\fi
\expandafter\ifx\csname bibfnamefont\endcsname\relax
  \def\bibfnamefont#1{#1}\fi
\expandafter\ifx\csname citenamefont\endcsname\relax
  \def\citenamefont#1{#1}\fi
\expandafter\ifx\csname url\endcsname\relax
  \def\url#1{\texttt{#1}}\fi
\expandafter\ifx\csname
urlprefix\endcsname\relax\def\urlprefix{URL }\fi
\providecommand{\bibinfo}[2]{#2}
\providecommand{\eprint}[2][]{\url{#2}}

\bibitem[{\citenamefont{Tsukada et~al.}(1999)\citenamefont{Tsukada, Sasago,
  Uchinokura, Zheludev, Maslov, Shirane, Kakurai, and Ressouche}}]{Tsukada99}
\bibinfo{author}{\bibfnamefont{I.}~\bibnamefont{Tsukada}},
  \bibinfo{author}{\bibfnamefont{Y.}~\bibnamefont{Sasago}},
  \bibinfo{author}{\bibfnamefont{K.}~\bibnamefont{Uchinokura}},
  \bibinfo{author}{\bibfnamefont{A.}~\bibnamefont{Zheludev}},
  \bibinfo{author}{\bibfnamefont{S.}~\bibnamefont{Maslov}},
  \bibinfo{author}{\bibfnamefont{G.}~\bibnamefont{Shirane}},
  \bibinfo{author}{\bibfnamefont{K.}~\bibnamefont{Kakurai}}, \bibnamefont{and}
  \bibinfo{author}{\bibfnamefont{E.}~\bibnamefont{Ressouche}},
  \bibinfo{journal}{Phys. Rev. B} \textbf{\bibinfo{volume}{60}},
  \bibinfo{pages}{6601} (\bibinfo{year}{1999}).

\bibitem[{\citenamefont{Zheludev et~al.}(2003)\citenamefont{Zheludev, Raymond,
  Regnault, Essler, Kakurai, Masuda, and Uchinokura}}]{Zheludev2003-2}
\bibinfo{author}{\bibfnamefont{A.}~\bibnamefont{Zheludev}},
  \bibinfo{author}{\bibfnamefont{S.}~\bibnamefont{Raymond}},
  \bibinfo{author}{\bibfnamefont{L.-P.} \bibnamefont{Regnault}},
  \bibinfo{author}{\bibfnamefont{F.~H.~L.} \bibnamefont{Essler}},
  \bibinfo{author}{\bibfnamefont{K.}~\bibnamefont{Kakurai}},
  \bibinfo{author}{\bibfnamefont{T.}~\bibnamefont{Masuda}}, \bibnamefont{and}
  \bibinfo{author}{\bibfnamefont{K.}~\bibnamefont{Uchinokura}},
  \bibinfo{journal}{Phys. Rev. B} \textbf{\bibinfo{volume}{67}},
  \bibinfo{pages}{134406} (\bibinfo{year}{2003}).

\bibitem[{\citenamefont{Zheludev et~al.}(2001)\citenamefont{Zheludev,
  Kenzelmann, Raymond, Masuda, Uchinokura, and Lee}}]{Zheludev2001-3}
\bibinfo{author}{\bibfnamefont{A.}~\bibnamefont{Zheludev}},
  \bibinfo{author}{\bibfnamefont{M.}~\bibnamefont{Kenzelmann}},
  \bibinfo{author}{\bibfnamefont{S.}~\bibnamefont{Raymond}},
  \bibinfo{author}{\bibfnamefont{T.}~\bibnamefont{Masuda}},
  \bibinfo{author}{\bibfnamefont{K.}~\bibnamefont{Uchinokura}},
  \bibnamefont{and} \bibinfo{author}{\bibfnamefont{S.-H.} \bibnamefont{Lee}},
  \bibinfo{journal}{Phys. Rev. B} \textbf{\bibinfo{volume}{65}},
  \bibinfo{pages}{014402} (\bibinfo{year}{2001}).

\bibitem[{\citenamefont{Zheludev et~al.}(2002)\citenamefont{Zheludev, kakurai,
  Masuda, Uchinokura, and Nakajima}}]{Zheludev2002-2}
\bibinfo{author}{\bibfnamefont{A.}~\bibnamefont{Zheludev}},
  \bibinfo{author}{\bibfnamefont{K.}~\bibnamefont{Kakurai}},
  \bibinfo{author}{\bibfnamefont{T.}~\bibnamefont{Masuda}},
  \bibinfo{author}{\bibfnamefont{K.}~\bibnamefont{Uchinokura}},
  \bibnamefont{and} \bibinfo{author}{\bibfnamefont{K.}~\bibnamefont{Nakajima}},
  \bibinfo{journal}{Phys. Rev. Lett.} \textbf{\bibinfo{volume}{89}},
  \bibinfo{pages}{197205} (\bibinfo{year}{2002}).

\bibitem[{\citenamefont{Zheludev et~al.}(2000)\citenamefont{Zheludev,
  Kenzelmann, Raymond, Ressouche, Masuda, kakurai, Maslov, Tsukada, Uchinokura,
  and Wildes}}]{Zheludev2000-2}
\bibinfo{author}{\bibfnamefont{A.}~\bibnamefont{Zheludev}},
  \bibinfo{author}{\bibfnamefont{M.}~\bibnamefont{Kenzelmann}},
  \bibinfo{author}{\bibfnamefont{S.}~\bibnamefont{Raymond}},
  \bibinfo{author}{\bibfnamefont{E.}~\bibnamefont{Ressouche}},
  \bibinfo{author}{\bibfnamefont{T.}~\bibnamefont{Masuda}},
  \bibinfo{author}{\bibfnamefont{K.}~\bibnamefont{kakurai}},
  \bibinfo{author}{\bibfnamefont{S.}~\bibnamefont{Maslov}},
  \bibinfo{author}{\bibfnamefont{I.}~\bibnamefont{Tsukada}},
  \bibinfo{author}{\bibfnamefont{K.}~\bibnamefont{Uchinokura}},
  \bibnamefont{and} \bibinfo{author}{\bibfnamefont{A.}~\bibnamefont{Wildes}},
  \bibinfo{journal}{Phys. Rev. Lett.} \textbf{\bibinfo{volume}{85}},
  \bibinfo{pages}{4799} (\bibinfo{year}{2000}).

\bibitem[{\citenamefont{Masuda et~al.}(2004)\citenamefont{Masuda, Zheludev,
  Uchinokura, Chung, and Park}}]{Masuda2004}
\bibinfo{author}{\bibfnamefont{T.}~\bibnamefont{Masuda}},
  \bibinfo{author}{\bibfnamefont{A.}~\bibnamefont{Zheludev}},
  \bibinfo{author}{\bibfnamefont{K.}~\bibnamefont{Uchinokura}},
  \bibinfo{author}{\bibfnamefont{J.-H.} \bibnamefont{Chung}}, \bibnamefont{and}
  \bibinfo{author}{\bibfnamefont{S.}~\bibnamefont{Park}},
  \bibinfo{journal}{Phys. Rev. Lett.} \textbf{\bibinfo{volume}{93}},
  \bibinfo{pages}{077206} (\bibinfo{year}{2004}).

\bibitem[{\citenamefont{Yamada et~al.}(2001{\natexlab{a}})\citenamefont{Yamada,
  Takano, and Hiroi}}]{Yamada2001}
\bibinfo{author}{\bibfnamefont{T.}~\bibnamefont{Yamada}},
  \bibinfo{author}{\bibfnamefont{M.}~\bibnamefont{Takano}}, \bibnamefont{and}
  \bibinfo{author}{\bibfnamefont{Z.}~\bibnamefont{Hiroi}}, \bibinfo{journal}{J.
  Alloys Compd.} \textbf{\bibinfo{volume}{317--318}}, \bibinfo{pages}{171}
  (\bibinfo{year}{2001}{\natexlab{a}}).

\bibitem[{\citenamefont{Yamada et~al.}(2001{\natexlab{b}})\citenamefont{Yamada,
  Hiroi, and Takano}}]{Yamada2001-2}
\bibinfo{author}{\bibfnamefont{T.}~\bibnamefont{Yamada}},
  \bibinfo{author}{\bibfnamefont{Z.}~\bibnamefont{Hiroi}}, \bibnamefont{and}
  \bibinfo{author}{\bibfnamefont{M.}~\bibnamefont{Takano}},
  \bibinfo{journal}{J. Solid State Chem.} \textbf{\bibinfo{volume}{156}},
  \bibinfo{pages}{101} (\bibinfo{year}{2001}{\natexlab{b}}).

\bibitem[{\citenamefont{Doty and Fischer}(1992)}]{Doty1992}
\bibinfo{author}{\bibfnamefont{C.~A.} \bibnamefont{Doty}} \bibnamefont{and}
  \bibinfo{author}{\bibfnamefont{D.~S.} \bibnamefont{Fischer}},
  \bibinfo{journal}{Phys. Rev. B} \textbf{\bibinfo{volume}{45}},
  \bibinfo{pages}{2167} (\bibinfo{year}{1992}).

\bibitem[{\citenamefont{Dasgupta and Ma}(1980)}]{Dasgupta1980}
\bibinfo{author}{\bibfnamefont{C.}~\bibnamefont{Dasgupta}} \bibnamefont{and}
  \bibinfo{author}{\bibfnamefont{S.~K.} \bibnamefont{Ma}},
  \bibinfo{journal}{Phys. Rev. B} \textbf{\bibinfo{volume}{22}},
  \bibinfo{pages}{1305} (\bibinfo{year}{1980}).

\bibitem{Masuda2006-erratum} T. Masuda, A. Zheludev, K. Uchinokura,
J.-H. Chung, S. Park,  Phys. Rev. Lett. 96, 169908 (2006).


\bibitem[{\citenamefont{Dender}(1997)}]{Dender1997}
\bibinfo{author}{\bibfnamefont{D.~C.} \bibnamefont{Dender}}, Ph.D. thesis,
  \bibinfo{school}{Johns Hopkins University} (\bibinfo{year}{1997}).

\bibitem[{\citenamefont{Lake et~al.}(2005)\citenamefont{Lake, Tennant, Frost,
  and Nagler}}]{Lake2005}
\bibinfo{author}{\bibfnamefont{B.}~\bibnamefont{Lake}},
  \bibinfo{author}{\bibfnamefont{D.~A.} \bibnamefont{Tennant}},
  \bibinfo{author}{\bibfnamefont{C.~D.} \bibnamefont{Frost}}, \bibnamefont{and}
  \bibinfo{author}{\bibfnamefont{S.~E.} \bibnamefont{Nagler}},
  \bibinfo{journal}{Nature Materials} \textbf{\bibinfo{volume}{4}},
  \bibinfo{pages}{329} (\bibinfo{year}{2005}).

\bibitem[{\citenamefont{Cloizeau and Pearson}(1962)}]{DesCloizeau1962}
\bibinfo{author}{\bibfnamefont{J.~D.} \bibnamefont{Cloizeau}} \bibnamefont{and}
  \bibinfo{author}{\bibfnamefont{J.~J.} \bibnamefont{Pearson}},
  \bibinfo{journal}{Phys. Rev} \textbf{\bibinfo{volume}{128}},
  \bibinfo{pages}{2131} (\bibinfo{year}{1962}).

\bibitem[{Mas()}]{Masudaunpublished}
\bibinfo{note}{T. Masuda, unpublished (2002)}.

\bibitem[{\citenamefont{Essler et~al.}(1997)\citenamefont{Essler, Tsvelik, and
  Delfino}}]{Essler1997}
\bibinfo{author}{\bibfnamefont{F.~H.~L.} \bibnamefont{Essler}},
  \bibinfo{author}{\bibfnamefont{A.~M.} \bibnamefont{Tsvelik}},
  \bibnamefont{and} \bibinfo{author}{\bibfnamefont{G.}~\bibnamefont{Delfino}},
  \bibinfo{journal}{Phys. Rev. B} \textbf{\bibinfo{volume}{56}},
  \bibinfo{pages}{11001} (\bibinfo{year}{1997}).

\bibitem[{\citenamefont{Dender et~al.}(1996)\citenamefont{Dender, davidovic,
  Reich, and Broholm}}]{Dender1996}
\bibinfo{author}{\bibfnamefont{D.~C.} \bibnamefont{Dender}},
  \bibinfo{author}{\bibfnamefont{D.}~\bibnamefont{davidovic}},
  \bibinfo{author}{\bibfnamefont{D.~H.} \bibnamefont{Reich}}, \bibnamefont{and}
  \bibinfo{author}{\bibfnamefont{C.}~\bibnamefont{Broholm}},
  \bibinfo{journal}{Phys. Rev. B} \textbf{\bibinfo{volume}{53}},
  \bibinfo{pages}{2583} (\bibinfo{year}{1996}).

\bibitem[{\citenamefont{Tennant
  et~al.}(1995{\natexlab{a}})\citenamefont{Tennant, Cowley, Nagler, and
  Tsvelik}}]{Tennant1995}
\bibinfo{author}{\bibfnamefont{D.~A.} \bibnamefont{Tennant}},
  \bibinfo{author}{\bibfnamefont{R.~A.} \bibnamefont{Cowley}},
  \bibinfo{author}{\bibfnamefont{S.~E.} \bibnamefont{Nagler}},
  \bibnamefont{and} \bibinfo{author}{\bibfnamefont{A.~M.}
  \bibnamefont{Tsvelik}}, \bibinfo{journal}{Phys. Rev. B}
  \textbf{\bibinfo{volume}{52}}, \bibinfo{pages}{13368}
  (\bibinfo{year}{1995}{\natexlab{a}}).

\bibitem[{\citenamefont{Tennant
  et~al.}(1995{\natexlab{b}})\citenamefont{Tennant, Nagler, nd~G.~Shirane, and
  Yamada}}]{Tennant1995-2}
\bibinfo{author}{\bibfnamefont{D.~A.} \bibnamefont{Tennant}},
  \bibinfo{author}{\bibfnamefont{S.~E.} \bibnamefont{Nagler}},
  \bibinfo{author}{\bibfnamefont{D.~W.} \bibnamefont{nd~G.~Shirane}},
  \bibnamefont{and} \bibinfo{author}{\bibfnamefont{K.}~\bibnamefont{Yamada}},
  \bibinfo{journal}{Phys. Rev. B} \textbf{\bibinfo{volume}{52}},
  \bibinfo{pages}{13381} (\bibinfo{year}{1995}{\natexlab{b}}).

\bibitem[{\citenamefont{Zaliznyak et~al.}(2004)\citenamefont{Zaliznyak, Woo,
  Perring, Broholm, Frost, , and Takagi}}]{Zaliznyak2004}
\bibinfo{author}{\bibfnamefont{I.~A.} \bibnamefont{Zaliznyak}},
  \bibinfo{author}{\bibfnamefont{H.}~\bibnamefont{Woo}},
  \bibinfo{author}{\bibfnamefont{T.~G.} \bibnamefont{Perring}},
  \bibinfo{author}{\bibfnamefont{C.~L.} \bibnamefont{Broholm}},
  \bibinfo{author}{\bibfnamefont{C.~D.} \bibnamefont{Frost}}, ,
  \bibnamefont{and} \bibinfo{author}{\bibfnamefont{H.}~\bibnamefont{Takagi}},
  \bibinfo{journal}{Phys. Rev. Lett.} \textbf{\bibinfo{volume}{93}},
  \bibinfo{pages}{087202} (\bibinfo{year}{2004}).

\bibitem[{\citenamefont{Muller et~al.}(1981)\citenamefont{Muller, Thomas, Puga,
  and Beck}}]{Muller81}
\bibinfo{author}{\bibfnamefont{G.}~\bibnamefont{Muller}},
  \bibinfo{author}{\bibfnamefont{H.}~\bibnamefont{Thomas}},
  \bibinfo{author}{\bibfnamefont{M.~W.} \bibnamefont{Puga}}, \bibnamefont{and}
  \bibinfo{author}{\bibfnamefont{H.}~\bibnamefont{Beck}}, \bibinfo{journal}{J.
  Phys. C} \textbf{\bibinfo{volume}{14}}, \bibinfo{pages}{3399}
  (\bibinfo{year}{1981}).

\bibitem[{\citenamefont{Schulz}(1986)}]{Schulz86}
\bibinfo{author}{\bibfnamefont{H.~J.} \bibnamefont{Schulz}},
  \bibinfo{journal}{Phys. Rev. B} \textbf{\bibinfo{volume}{34}},
  \bibinfo{pages}{6372} (\bibinfo{year}{1986}).

\bibitem[{Smi()}]{Smirnov_unpublished}
\bibinfo{note}{A. I. Smirnov,- unpublished (2005).}

\bibitem[{Ehl()}]{Ehlers_unpublished}
\bibinfo{note}{G. Ehlers, A. Zheludev and J. Gardener,- unpublished (2006).}

\end{thebibliography}
\end{document}